\documentclass[usenatbib]{mn2e}
\usepackage{epsfig}
\usepackage{times}
\title[The H~{\sc i}, He~{\sc i} and He~{\sc ii} free-bound continuous emission spectra] {Theoretical calculations of the H~{\sc i}, He~{\sc i} and He~{\sc ii} free-bound continuous emission spectra}\author[Ercolano \& Storey]{B. Ercolano$^1$, P. J. Storey$^1$\\
$^1$Department of Physics and Astronomy, University College London, Gower Street, London WC1E~6BT, UK\\}
\date{Received:}

\begin{document}
\setcounter{table}{1}

\maketitle

\begin{abstract}

We present coefficients for the calculation of the continuous emission
spectra of H~{\sc i}, He~{\sc i} and He~{\sc ii} due to electron-ion
recombination. Coefficients are given for photon energies from the
first ionization threshold for each ion to the $n=20$ threshold of
hydrogen ($36.5\mu$), and for temperatures
100~K$~\le~T~\le~10^5$~K. The emission coefficients for He~{\sc i} are
derived from accurate {\it ab initio} photoionization data. The
coefficients are scaled in such a way that they may be interpolated by a
simple scheme with uncertainties less than 1\% in the whole
temperature and wavelength domain. The data are suitable for
incorporation into photoionisation/plasma codes and should aid with
the interpretation of spectra from the very cold ionised gas phase
inferred to exist in a number of gaseous clouds.  

\end{abstract}

\nokeywords


\section{Introduction}

Analyses of optical recombination line (ORL) spectra of photoionised
regions have suggested the existence of cold ionised gas
(100~K--2000K), mixed within a warmer component at more typical
nebular electron temperatures (8000~K--10000~K).  The pockets of cold
ionised gas of high metal content are invoked as a possible
explanation of the long-standing problem of the discrepancy between
elemental abundances derived from ORLs and those derived from
collisionally excited lines (CELs).  In this scenario, ORLs and CELs
are preferentially emitted by the cold and warm phases, respectively
(for a recent review see Liu, 2002); this problem is closely linked to
the observation that Balmer jump temperatures of H~{\sc ii} regions and Planetary Nebulae are
systematically lower than those derived from the $[$O~{\sc iii}$]$
nebular to Auroral line ratio (Liu \& Danziger 1993).

Understanding the complex spectra arising from such regions relies on
the construction of detailed photoionisation models able to account
for all gas phases that may be present. Currently, one of the major
limitations in such modelling is the lack of an accurate atomic data
set extending to such low temperatures. The need of low-temperature
effective recombination coefficient for the calculation of
recombination lines for metals is apparent. 
However, in addition to the discrete emission line spectrum, a
continuous emission is also produced by the ionised gas, mainly due to
free-bound recombination processes of hydrogen and helium ions,
free-free transitions in the Coulomb fields of H$^+$, He$^+$ and
He$^{2+}$ and two-photon decay of the 2~$^2$S$_{1/2}$ level of H~{\sc
i} and He~{\sc ii}, and, less importantly of the 2~$^1$S$_{0}$ level
of He~{\sc i}. Accurate continuous emission coefficients are essential
for the correct prediction of the Balmer jump by photoionisation
codes.

The importance of the continuum processes listed above has long been known and
emission coefficients have been tabulated (see e.g. Seaton 1955, 1960;
Brown \& Mathews 1970), for a range of temperatures and wavelengths
mainly aimed at the study of optical data for classical H~{\sc ii}
regions. Ferland (1980) derived H~{\sc i} and He~{\sc ii} continuous
emission and recombination coefficients for a wider range of
temperatures (500~K--2$\cdot$10$^6$~K) and wavelengths to aid the
interpretation of ultraviolet (UV) and infrared (IR) as well as
optical data from nova ejecta.

In this work we present new calculations of the H~{\sc i}, He~{\sc i}
and He~{\sc ii} continuous emission coefficients due to free-bound
recombination, over temperatures ranging from 100~K to 10$^{5}$~K, for
the physical conditions thought likely to occur in chemically
inhomogeneous regions, which may include pockets of cold ionised material
intermixed within typical nebular gas.

\section{Calculations Method}

   \begin{figure*}
   \centering
   \begin{minipage}[t]{8.5cm}
     \includegraphics[width=8.5cm]{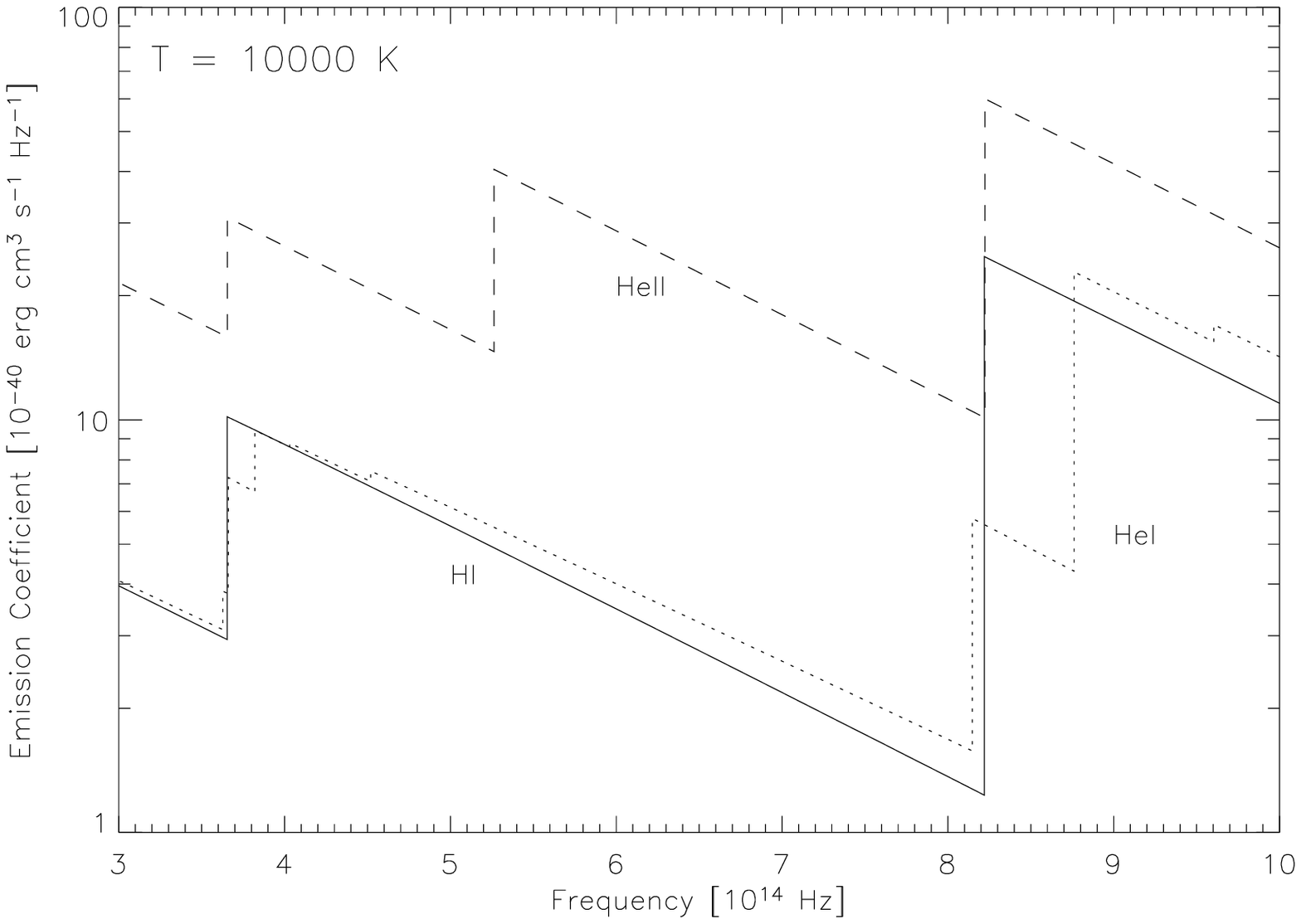}
   \end{minipage}
   \begin{minipage}[t]{8.5cm}
     \includegraphics[width=8.5cm]{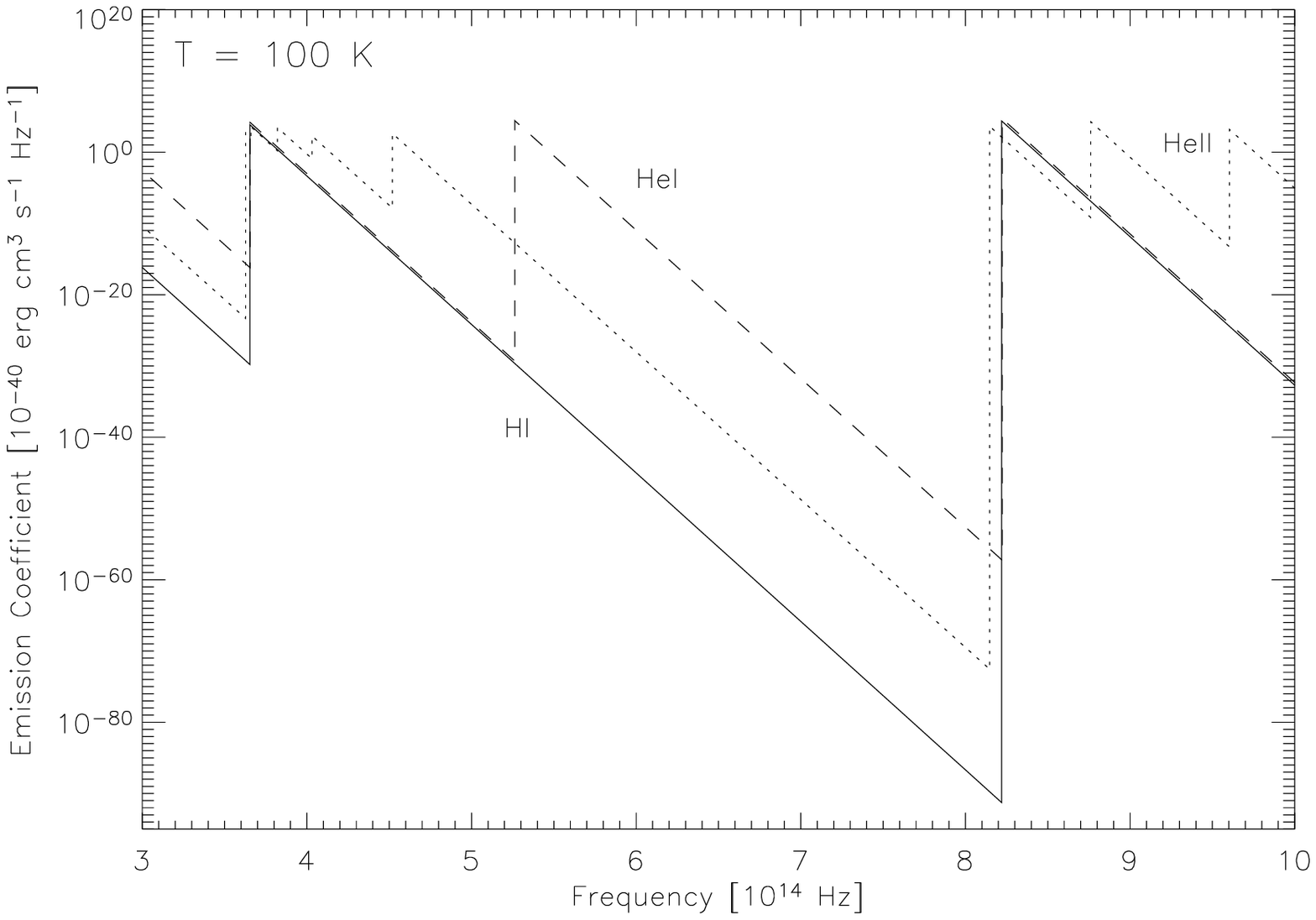}
   \end{minipage}
   \caption{Frequency-dependent continuum emission coefficients $\gamma(\nu)$ for H~{\sc i} ({\it solid line}), $\gamma(\nu)$ for He~{\sc i} ({\it dotted line}) and $\gamma(\nu)$ for He~{\sc ii} ({\it dashed line}), calculated at T~=~10000~K (left panel, to be compared with Figure~1 from Brown \& Matthews, 1970) and  T~=~100~K (right panel).
   }
         \label{fig1}
   \end{figure*}

Using the Saha-Boltzmann equation and the Milne relation, we may
express the continuous emission coefficient $\gamma(\nu)$
corresponding to the recombination process
$$ X^+ + {\rm e}^-(\epsilon) \rightarrow X^* +  h\nu $$
in terms of the photoionization cross-section $\sigma_{\nu}(X^*)$ 
$$ \gamma(\nu) = \frac{4\pi h}{c^2} (\frac{h^2}{2\pi m k T})^{3/2} {\rm e}^{-\epsilon/kT} \frac{\omega^*}{\omega^+} \nu^3 \sigma_{\nu}(X^*) $$ 
where $\omega^+$ and
$\omega^*$ are the statistical weights of the recombining ion initial
state and final state respectively, and $\epsilon$ is the free
electron energy. In terms of $\gamma(\nu)$, the energy emitted
per unit volume per unit time in frequency interval $\nu$ to
$\nu + {\rm d\nu}$ is $N_e N(X^+) \gamma(\nu) {\rm d}\nu$.

For the contribution $\gamma_n(\nu)$ at frequency $\nu$ from
recombinations to states of principal quantum number $n$ of H$^0$ and
He$^+$ we compute the necessary energy-dependent photoionization
cross-sections using the hydrogenic codes described by Storey \&
Hummer (1991). The total emission coefficient $\gamma(\nu)$ is then
$$ \gamma(\nu) = \sum_{n_0}^{\infty} \gamma_n(\nu) $$ where in
practice we truncate the sum at $n=200$ for H~{\sc i} and $n=350$ for
He~{\sc ii}, which is sufficient to ensure convergence to all figures
given in the Tables.

For recombination to atomic helium we use the {\it ab initio}
calculated photoionization cross-sections described by Hummer \&
Storey (1998). Cross-section data are available for $n^1$S, $n^3$S,
$n^1$P$^{\rm o}$, $n^3$P$^{\rm o}$, $n^1$D, $n^3$D, $n^1$F$^{\rm o}$
and $n^3$F$^{\rm o}$states with $l+1\leq n \leq 20$. For higher values
of $n$ and for $l>3$ we use hydrogenic data. Thus for $n\geq 5$ the
emission coefficient has nine distinct thresholds for each $n$
corresponding to the eight separate terms plus a threshold at the
hydrogenic energy. Hummer \& Storey (1998) showed that the results of
their {\it ab initio} calculation of the photoionization cross-sections
are in better agreement at threshold with the highly accurate
bound-bound calculations of Drake (1996) than any of the other methods
used to compute helium recombination processes.



Contributions due to bremsstrahlung emission of a Maxwellian
distribution of electrons in the Coulomb fields of hydrogen and helium
ions are not included in our results, but can be obtained readily using
(e.g.) equation~2 of Brown \& Mathews (1970) or the free-free
computer code published by Storey \& Hummer (1991).

Contributions to the continuum emission from two-photon emission are
also not included but can be computed using the formulae of Nussbaumer
\& Schmutz (1984).

\begin{table*}
\begin{minipage}[t][70mm]{\textwidth}
\centering
\center\caption{Representative values of the exact and interpolated coefficients, $\gamma(\nu)$}             
\begin{tabular}{ccccccccc}
\hline       
\noalign{\vskip3pt}
$\nu~[{\rm Hz}]$  & E $[{\rm Ryd}]$ & $\lambda_{\rm vac}[$\AA$]$ & \multicolumn{6}{c}{$\gamma(\nu) [10^{-40}$erg cm$^{3}$ s$^{-1}$ Hz$^{-1}]$}\\
\noalign{\vskip3pt}
                  &                 &                        &  H\,{\sc i}(exact) &  H\,{\sc i}(interp) &  He\,{\sc i}(exact) &  He\,{\sc i}(interp) &  He\,{\sc ii}(exact) &  He\,{\sc ii}(interp) \\
\noalign{\vskip3pt}
\hline       
\noalign{\vskip3pt}
   1.000($+$14)  &     0.030397   &   29979.246  &    2.415($+$01) &     2.418($+$01) &    2.408($+$01)  &    2.408($+$01) &    7.393($+$01)  &    7.412($+$01) \\
   2.000($+$14)  &     0.060793   &   14989.623  &    2.852($+$00) &     2.858($+$00) &    3.026($+$00)  &    3.028($+$00) &    3.222($+$01)  &    3.233($+$01) \\
   3.000($+$14)  &     0.091190   &    9993.082  &    1.653($+$00) &     1.655($+$00) &    2.067($+$00)  &    2.068($+$00) &    7.034($+$01)  &    7.047($+$01) \\
   4.000($+$14)  &     0.121586   &    7494.811  &    4.442($+$01) &     4.442($+$01) &    5.948($+$01)  &    5.947($+$01) &    9.356($+$01)  &    9.366($+$01) \\
   5.000($+$14)  &     0.151983   &    5995.849  &    6.996($-$01) &     6.988($-$01) &    2.848($+$00)  &    2.856($+$00) &    1.457($+$00)  &    1.458($+$00) \\
   6.000($+$14)  &     0.182380   &    4996.541  &    1.093($-$02) &     1.092($-$02) &    5.478($-$02)  &    5.495($-$02) &    3.098($+$01)  &    3.098($+$01) \\
   7.000($+$14)  &     0.212776   &    4282.749  &    1.698($-$04) &     1.694($-$04) &    1.015($-$03)  &    1.017($-$03) &    4.807($-$01)  &    4.805($-$01) \\
   8.000($+$14)  &     0.243173   &    3747.406  &    2.629($-$06) &     2.628($-$06) &    1.828($-$05)  &    1.831($-$05) &    7.437($-$03)  &    7.438($-$03) \\
   9.000($+$14)  &     0.273569   &    3331.027  &    2.376($+$01) &     2.376($+$00) &    1.795($+$02)  &    1.795($+$02) &    5.054($+$01)  &    5.054($+$01) \\
   1.000($+$15)  &     0.303966   &    2997.925  &    3.704($-$01) &     3.702($-$01) &    1.097($+$01)  &    1.100($+$01 &    7.823($-$01)  &    7.821($-$01) \\
\noalign{\vskip3pt}
\hline                  
\end{tabular}
\end{minipage}
\end{table*}



   \begin{figure*}
   \centering
   \includegraphics[width=20cm]{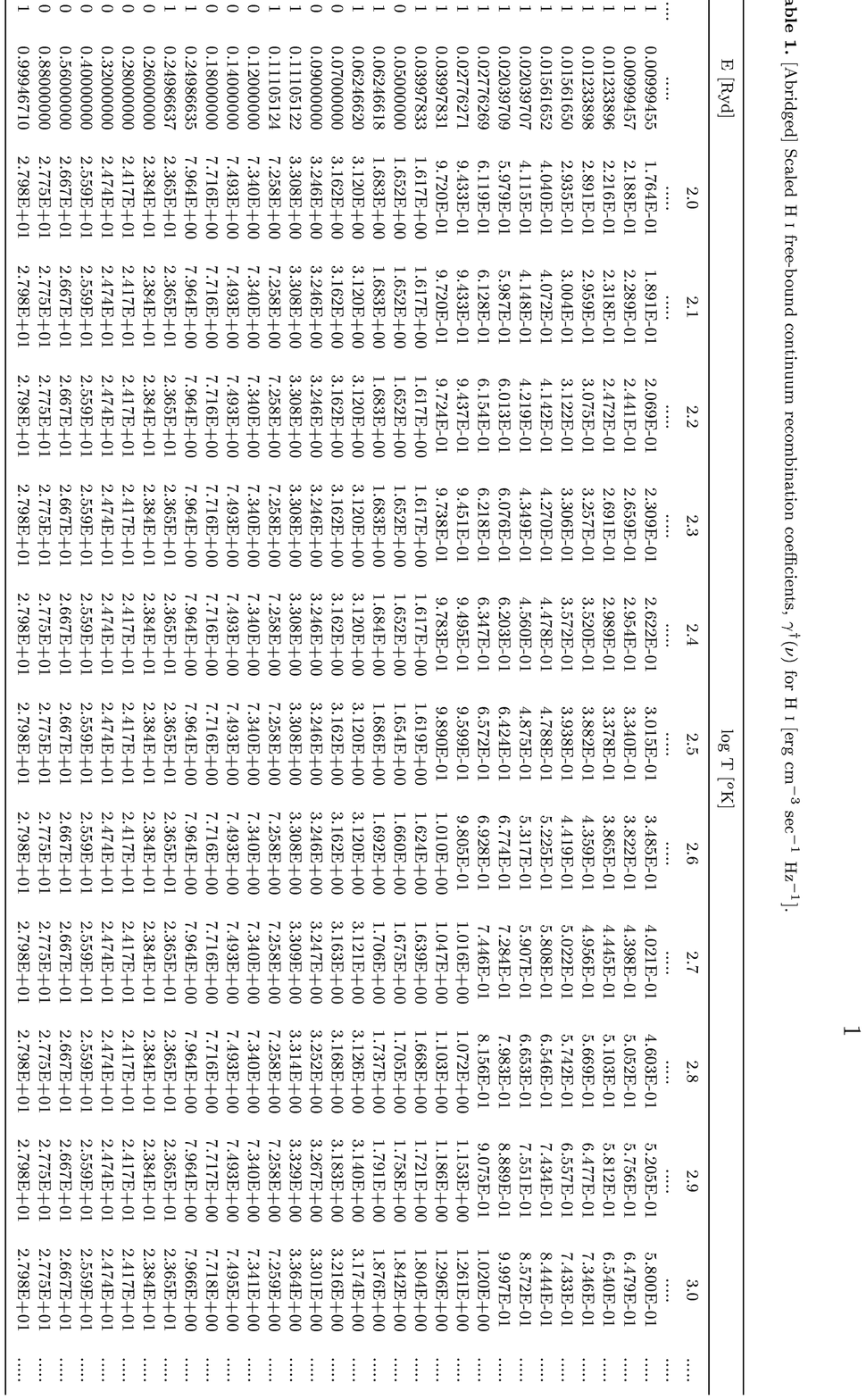}             
         \label{t}
   \end{figure*}

\section{Results}

 The full tables of continuous emission coefficients, Tables 3, 4 and
 5 for H~{\sc i}, He~{\sc ii} and He~{\sc i} respectively are
 available in electronic form only. In Table~1 we show an extract from
 the table for H~{\sc i}. Coefficients are tabulated for log~$T$[K] =
 2.0(0.1)5.0 and for photon energies (in Rydbergs) from just less than
 the ground state threshold energy to just above the energy of the
 threshold at $n=20$. Values are tabulated on either side of each
 threshold and at some additional nodal points inserted to make
 interpolation sufficiently accurate. At low temperatures the
 coefficients fall rapidly and exponentially at energies above each
 threshold making interpolation difficult so rather than $\gamma(\nu)$
 we tabulate $\gamma^{\dagger}(\nu)$ defined by

\begin{eqnarray}
\gamma^{\dagger}(\nu)&   = & \gamma(\nu)\ 10^{34}\  T^{3/2} {\rm e}^{{\Delta E}/kT} \\
                     &   = & \gamma(\nu)\ 10^{40}\ t^{3/2} {\rm e}^{{15.7887\Delta E_R}/t} 
\end{eqnarray}

\noindent where $t~=~T[K]/10^{4}$, $\Delta E$ is the difference between the photon energy, h$\nu$,
and the energy of the nearest threshold of lower
energy and $\Delta E_R$ is the same energy in Rydberg units. Thresholds are indicated by index 1 and additional nodal
points by index 0.

The recommended procedure for deriving the emission coefficient at a
given temperature and photon energy is to interpolate linearly in the
appropriate table in the variables log$~T$ and photon energy to obtain
the scaled coefficient $\gamma^{\dagger}(\nu)$.  Equation~1 can then
be applied to obtain $\gamma(\nu)$.  Table~2 gives exact values of
$\gamma(\nu)$ and values derived from the recommended interpolation
scheme for each ion at a range of photon energies.


Comparison of the tabulated values with calculations performed on a
finer frequency grid and temperature grid show that linear
interpolation in log~$T$ and photon energy yields accurate results
with maximum deviations of 1\% and average deviations much smaller
than this.  

Figure~1 shows the non-scaled continuum emission coefficients,
$\gamma(\nu)$ for H~{\sc i}, He~{\sc i} and He~{\sc ii} in the optical
wavelength range for a temperature of 10000~K (left panel) and 100~K
(right panel). The left panel of figure~1 is directly comparable to
figure~1 of Brown \& Mathews (1970). We can only compare the magnitude
of the discontinuity at each threshold directly with the results of
Brown \& Matthews (1970), since their results incorporate all
continuum processes while ours only deal with free-bound processes.
Comparing results for H~{\sc i} and He~{\sc ii} we find a maximum
difference of 1\% for H~{\sc i} at the Balmer threshold and at the
lowest temperature tabulated by them of 4000K.  The difference is
attributable to Brown \& Matthews using an approximate expression for
the hydrogenic threshold photoionization cross-sections while we use
the exact expressions incorporated in the codes of Storey \& Hummer
(1991). A similar comparison for He~{\sc i} shows that for all
thresholds except that corresponding to the 3~$^3$P$^{\rm o}$ state at
an air wavelength of 7849\AA, our results differ by no more than 2.1\%
from those of Brown \& Matthews (1970). The differences that do exist
are due to the approximate method used by Brown \& Matthews to
calculate the helium photoionization cross-sections. In the case of
the 3~$^3$P$^{\rm o}$ threshold, we find much larger differences reaching 21\%
at 4000K. This is almost certainly due to a numerical error in the
work of Brown \& Matthews, since the magnitude of the discontinuity at
the 3~$^3$P$^{\rm o}$ threshold does not obey the correct scaling with
temperature in their work.

We also compared our values of $\gamma(\nu)$ for H~{\sc i} and
He~{\sc ii} with those published by Ferland (1980) and found, in
general, good agreement, with typical deviations of the order of 2-5\%
in the overlapping temperature range.



The electronic tables are structured as follows: node/threshold (0/1)
indices are given in column~1, the photon energies in Ryd are given in
column~2, the scaled free-bound emission coefficients for temperatures
in the 100\,K-100000\,K range are given in the remaining columns. For
He~{\sc i} and He~{\sc ii} coefficients are tabulated for log~$T$[K] =
2.0(0.1)5.0 while for He~{\sc i}, log~$T$[K] = 2.0(0.04)5.0. The maximum
photon energy is slightly less than the ground state ionization energy
for each ion and the minimum photon energy corresponds to the $n=20$
threshold in H~{\sc i}.



\section{Conclusions}
We have presented new calculations of the free-bound continuous
emission coefficients for the hydrogen and helium ions. The results
for He~{\sc i} are derived from accurate {\it ab initio}
photoionization cross-section data. The coefficients 
are given for a wide range of temperatures and frequencies, extending
to the previously unexplored very low-temperatures regime. This is
needed for the interpretation of spectra from very cold ionised gas,
which has been inferred to exist, possibly in the form of
density/chemical inhomogeneities, from observations of ORLs in H~{\sc
ii} regions and planetary nebulae. The data are presented under an
interpolation scheme that allows estimates to be obtained over the
entire temperature and energy range presented with less than 1\%
uncertainty.


\bibliographystyle{mn2e}

\bibliography{references}

\end{document}